# Energy-Resolved Femtosecond Hot Electron Dynamics in Single Plasmonic Nanoparticles

Jacob Pettine,[1,2,3] * Paolo Maioli,[4] Fabrice Vallee,[4] Natalia Del Fatti,[4,5] and David J. Nesbitt[1,2,6] *

[1] JILA, University of Colorado Boulder and the National Institute of Standards and Technology, Boulder, Colorado 80309, United States

[2] Department of Physics, University of Colorado Boulder, Boulder, Colorado 80309, United States

[3] Center for Integrated Nanotechnologies, Los Alamos National Laboratory, Los Alamos, NM 87545, United States

[4] Université de Lyon, CNRS, Université Claude Bernard Lyon 1, Institut Lumière Matière, 69622 Villeurbanne Cedex, France

[5] Institut Universitaire de France (IUF), France

[6] Department of Chemistry, University of Colorado Boulder, Boulder, Colorado 80309, United States

*Correspondence may be addressed to J.P. (jacob.pettine@lanl.gov) or D.J.N. (djn@jila.colorado.edu)

# Abstract

Efficient excitation and harvesting of hot carriers from nanoscale metals is central to many emerging photochemical, photovoltaic, and ultrafast optoelectronic applications. Yet direct experimental evidence of the relevant femtosecond dynamics in ubiquitous tens-of-nanometer gold structures remains lacking, despite the rich interplay between interfacial and internal plasmonic fields, excitation distributions, and scattering processes. To explore the effects of nanoscale structure on these dynamics, we employ a new technique for simultaneous time-, angle-, and energy-resolved photoemission spectroscopy of single plasmonic nanoparticles. Photoelectron velocity distributions reveal bulk-like ballistic hot electron transport in nanorod and nanoshell geometries, with no evidence of surface effects. Energy-resolved dynamics observed in the 1–2 eV range and extrapolated to lower energies via kinetic Boltzmann theory provide the first direct measurements of hot carrier lifetimes within nanoscale gold. Remarkably, we find that particles with dimensions as small as 10 nm serve as exemplary platforms for studying intrinsic metal dynamics.

# Introduction

Plasmonic oscillations in metal nanoparticles generate strong, nanolocalized optical field enhancements and high densities of photoexcited electrons and holes. Whether nascent or internally thermalized via electron-electron scattering, such excitations are often referred to as "hot" charge carriers, with average optical excitation energies much greater than the thermal energy of the metal lattice, $k_B T_l$. Efficient harvesting of these hot carriers is critical for a broad range of emerging applications, including plasmon-enhanced photocatalysis (1-3), solar photovoltaics (4-6), biotherapeutics (7), and ultrafast integrated photodetection (8, 9). However, the ability to tailor nanoplasmonic systems for optimized hot carrier collection efficiencies remains an ongoing challenge, requiring detailed knowledge of energy-dependent femtosecond hot carrier lifetimes and their nanoscale geometry-dependent spatial and vector momentum distributions.

Energy-averaged hot carrier dynamics have been well studied in metal nanoparticles using ultrafast optical pump-probe measurements of the transient dielectric response (10-13). Such studies have been central to understanding picosecond electron-lattice thermalization and impulsive acoustic excitations (14-16), also offering insight into hundreds-of-femtosecond electron-electron thermalization times (17). However, the overall energy-averaged hot carrier response probed in these studies is disproportionately influenced by longer-timescale ($> 100$ fs), low-energy ($< 1$ eV) thermalization dynamics rather than the much faster tens-of-femtosecond decay times of the higher-energy nascent carriers. Energy-resolved studies are therefore needed to extract these faster dynamics, which are most relevant for charge emission over interfacial barriers (e.g., ~1.1 eV for an Au-TiO$_2$ interface (2, 4)) and transfer into discrete admolecular surface states.

Energy-resolved two-photon photoemission (2PPE) has been commonly utilized to probe hot carrier dynamics in bulk and thin film metal systems (18-20), while studies of femtosecond carrier lifetimes in metal nanoparticles remain relatively limited. Stimulated by initial investigations of nanoparticle ensembles (21-23), nanoscale photoemission capabilities have steadily evolved over the past couple decades, with nanometer-scale photoemission hot spots (24, 25), few-femtosecond plasmon dephasing dynamics (26, 27), energy-dependent lifetime trends (28, 29), and photoelectron momentum distributions (30-34) all measured at the single nanoparticle level. Complementary progress has been made in theoretically modeling photoelectron spatial and momentum distributions in arbitrary nanoscale geometries (32, 35-37). Despite these advances, the direct characterization of energy-dependent hot carrier lifetimes in nanostructured gold has been critically overlooked, while an even more pressing need for time- and angle-resolved dynamical insights into the correspondence between nanoscale geometry, spatial excitation density, and carrier momentum distributions has eluded experiment. Ideally, such studies would offer single-nanoparticle resolution to further elucidate which behaviors vary and which remain consistent from particle to particle, in the presence of defects, orientational effects, resonance shifts, and inter-particle coupling.

To help address these problems, we combine femtosecond pump-probe excitation, scanning photoemission microscopy, and velocity-resolved detection methods to view the femtosecond hot electron dynamics within single, well-characterized gold nanoparticles. The tens-of-nanometer gold morphologies examined here are used ubiquitously in a variety of plasmonics applications, lying within an intermediate nanoscale size range that precludes quantum confinement effects but still exhibits high (and highly tunable) surface-to-volume ratios. A detailed interplay thus naturally emerges between surface and bulk excitation and emission pathways, which can both contribute to energy flow but with different hot carrier spatial, temporal, and momentum distribution. The challenge is distinguishing these interfacial versus ballistic dynamical effects and energy-resolved population kinetics, which we can deconstruct with simultaneous time, angle, and energy resolution.

## Results

**Two-color photoemission from single nanoparticles**

The experimental configuration for time- and angle-resolved scanning photoemission spectroscopy is shown in Fig. 1, with further details provided in the Methods. To clarify the effects of shape, size, and crystallinity on hot electron dynamics, two distinctly different nanoparticle geometries are investigated: monocrystalline gold nanorods and polycrystalline gold/silica nanoshells (Fig. 2A). Small gold nanorods coated in cetyltrimethylammonium bromide (CTAB) stabilizing ligands with 10(1) nm diameter and 41(4) nm total tip-to-tip length (standard deviations in parenthesis) exhibit an average longitudinal dipolar surface plasmon resonance (SPR) around 710 nm when supported on an ITO substrate. For our much larger gold/silica nanoshells with lipoic acid stabilizing ligands, 160(7) nm overall diameter, and 120(4) nm silica core diameter, we observe an average dipolar SPR on ITO around 675 nm (30).

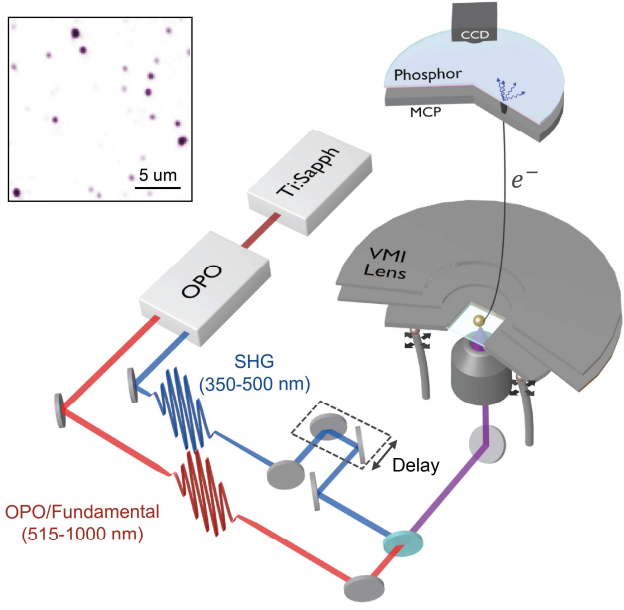

**Fig. 1. Time-resolved scanning photoemission spectroscopy.** Femtosecond dynamics in single nanoparticles measured via pump-probe excitation and velocity-resolved photoelectron mapping. Inset: Spatially-resolved scanning photoemission map (linear scale) of gold nanorods, with a range of photoemissivities due to different nanorod resonances and orientations relative to the linearly-polarized excitation light.

To overcome gold nanoparticle work functions ($> 4$ eV) while taking advantage of plasmonic enhancements for high signal contrast between nanoparticles and the ITO substrate, we perform two-color pump-probe measurements. Both nanorod and nanoshell samples are resonantly excited with a 1.77 eV (700 nm) pump photon, after which a non-resonant 3.1 eV probe photon ejects the hot electrons into vacuum. The $1 + 1'$ 2PPE process order is verified by the pump-probe signal intensity dependence (Figs. 2B and 2C), holding one beam intensity fixed while varying the other. To leading order in each process, the overall photoemission rate is

$$\Gamma_{\text{PE}} = \sigma_{\text{pp}}^{(1+1')} I_{\text{pump}} I_{\text{probe}} + \sigma_{\text{pump}}^{(3)} I_{\text{pump}}^3 + \sigma_{\text{probe}}^{(2)} I_{\text{probe}}^2 + \Gamma_{\text{ITO}}, \qquad (1)$$

in which the first term is the pump-probe signal of interest, while the remaining terms are weaker contributions from single-color pump 3PPE, probe 2PPE, and corresponding (weaker still) contributions from the ITO substrate. The pump-probe photoemission cross-section, $\sigma_{\text{pp}}^{(1+1')}$, and the single-color photoemission cross-sections, $\sigma_{\text{pump}}^{(3)}$ and $\sigma_{\text{probe}}^{(2)}$, are determined experimentally for each nanoparticle under pump-probe, pump-only, and probe-only exposure. The pump-probe contribution to the total signal is then optimized under $I_{\text{pump}}$ conditions for which $\sigma_{\text{probe}}^{(2)} I_{\text{probe}}^2 \approx \sigma_{\text{pump}}^{(3)} I_{\text{pump}}^3$. For the nanorods, the various single- and multi-photon cross-sections depend not only on the excitation frequencies, but also strongly on the linear polarization angles of both pump and probe beams, as demonstrated in Fig. 2D for pump 3PPE ($\cos^6 \theta$) and probe 2PPE ($\sim \cos^3 \theta$; with the non-resonant interaction deviating from simple $\cos^{2n} \theta$ dependence). This strong polarization dependence is due to the stronger electric field enhancements for the resonant pump beam as well as for the non-resonant probe beam when the incident field is aligned with the longitudinal nanorod axis. By contrast, the nanoshell photoemission signal displays no polarization dependence, as expected for the azimuthal symmetry of the supported spherical nanoshell in the absence of surface defect hot spots (30, 31). Despite the several hundred-fold larger area excited on the ITO substrate compared with the geometrical nanorod cross section, the nanorod pump-probe signal is still an order of magnitude stronger than the ITO contribution due to the plasmon-enhanced pump (and even enhanced off-resonant probe) absorption cross-sections.

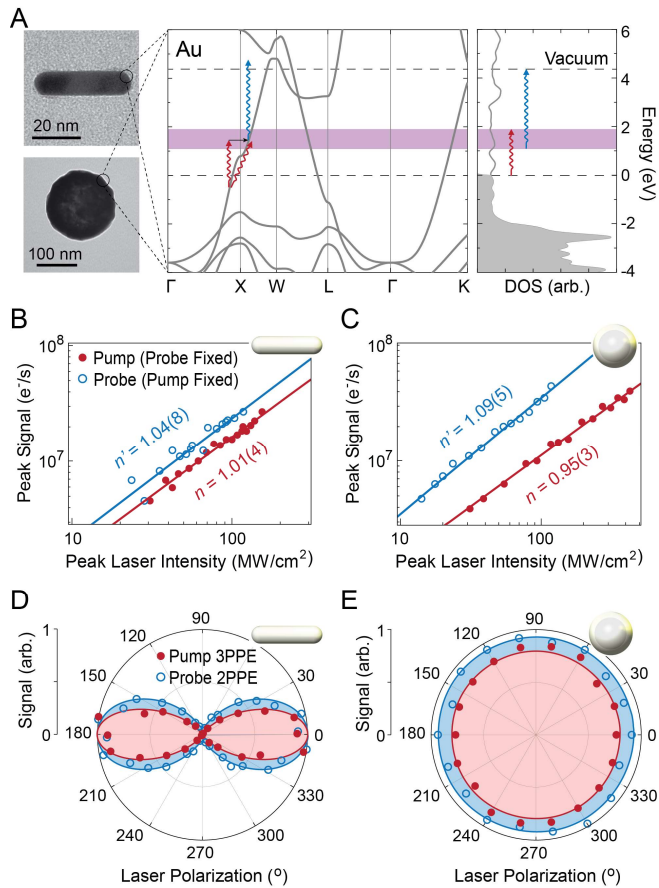

**Fig. 2. Two-color resonant + non-resonant pump-probe photoemission from gold nanorods and nanoshells.** (**A**) Transmission electron micrographs of a representative gold nanorod and gold nanoshell, shown alongside the electronic band structure and density of states of gold (Supplementary Information). For visible pump photon energies, phonon/defect-mediated and/or field-gradient-assisted intraband transitions are the dominant photoexcitation mechanisms leading to intermediate state (hot carrier) population. The purple zone indicates the overlap between the excitation and probing ranges for a 1.77 eV pump, 3.1 eV probe, and ~4.2 eV nanoparticle work functions. (**B**) Nanorod and (**C**) nanoshell laser intensity-dependent pump-probe photoemission signal, with one beam held fixed at an intermediate intensity while varying the other. Solid lines are power-law fits, verifying the $n + n' = 1 + 1'$ two-color 2PPE process in each case. (**D**) Nanorod and (**E**) nanoshell polarization-dependent single-color nonlinear photoemission signal, illustrating the strong nanorod longitudinal coupling for both pump and probe beams, compared with the polarization-insensitive nanoshell excitation. Solid lines are from finite element calculations (Methods), with the simulated plasmonic field enhancements, $|E/E_0|^{2n}$ (for single-color $n$-photon photoemission), integrated over the respective gold volumes.

**Geometry-insensitive carrier dynamics**

Nanoplasmonic photoexcitation and transfer/emission can occur directly at the surface or ballistically from within the bulk (32, 38-41). Direct surface excitations into hybridized metal-admolecular energy states (38, 39, 42-45), across metal-semiconductor Schottky barriers (5, 46), or over metal-vacuum barriers (47-49) occur predominantly within plasmonic surface hot spot regions and effectively bypass hot carrier dynamics within the metal. Time-resolved studies of such systems may thus reflect the ultrafast interfacial state dynamics (42, 46). For gold nanorods and nanoshells, however, it has been shown that resonant photoexcitation occurs predominantly in field-enhanced regions within the volume of the metal nanoparticles (31, 32), involving visible-frequency intraband transitions with momentum conserved either by phonon/defect scattering or field-gradient-assisted damping (Fig. 2A).The latter is due to large momentum components of rapidly spatially-varying plasmonic fields, which are known to occur in surface field regions of plasmonic particles or dimers (50) but have also recently been shown to significantly enhance volume intraband photoluminescence in gold nanorods (51). With the three-fold smaller nanorod diameters and even more dramatic internal field variations in the present studies (Fig. S8), significant volume excitation density from internal field-gradient-assisted damping may be expected. Other momentum-conserving excitation pathways, including intraband excitation from electron-electron scattering (52) and interband absorption from $5d$-band initial states (~2 eV below the Fermi level), yield lower-energy carriers that are not probed in the present studies. For the pathways that can contribute to final electron states above the vacuum level, transport of the photoexcited electrons to the interface can be treated as a three-step ballistic process (53): (i) photoexcitation within the metal, (ii) transport to an interface with quasi-elastic (e.g., electron-phonon) and inelastic (electron-electron) scattering along the way, and (iii) possible photoinjection into the surrounding medium given sufficient surface-normal momentum. While phenomenological, this semi-classical model captures the same essential features of the dynamics as the more rigorous one-step quantum theory of excitation from Bloch states into damped, inverse low-energy electron diffraction final states (54, 55).

Energy-integrated response functions of the ballistic hot electron dynamics are determined via cross-correlation measurements (Figs. 3A and 3B). The lower end of the accessible $E - E_\text{F} \approx 1.1–1.8$ eV intermediate state excitation energy range is constrained by the difference between the probe photon energy (3.1 eV) and the gold nanoparticle work function (~4.15 eV), while the upper end is constrained by the pump photon energy (1.77 eV). Hot electron dynamics are extracted from pump-probe time delay traces via the rate equation for the intermediate ($|m\rangle$) state population (22),

$$\dot{N}_m(t) = \frac{\sigma_{im} I_{\text{pump}}(t)}{\hbar \omega_{\text{pump}}} - \frac{1}{\tau_{ee}(E_m)} N_m(t), \qquad (2)$$

where $\sigma_{im}$ is the absorption cross-section for the initial state ($|i\rangle$) to $|m\rangle$ transition and $\tau_{ee}(E_m)$ is the hot electron lifetime ($T_1$ population decay time) at $E_m$. This simple limiting form of the three-level optical Bloch equations (18, 56) is well established in the case of rapid dephasing of any coherences relative to population times, as expected for continuum transitions (19, 22, 57). The solution of Eq. 2 is determined via bilateral Laplace transform, yielding the convolution $N_m(t) \propto \int_{-\infty}^{\infty} I_{\text{pump}}(t - \tau) R_e(\tau) d\tau$, in which $R_e(\tau)$ is the hot electron impulse response, given by $\Theta(\tau) e^{-\tau/\tau_{ee}(E_m)}$, where $\Theta$ is the Heaviside function. The photoemission rate as a function of time delay, $\dot{N}_f(t)$, is then proportional to the cross-correlation of $N_m(t)$ with $I_{\text{probe}}(t)$. It should be noted that while reverse decay contributions from the probe-first ($1' + 1$) excitation pathway can also occur, these contributions are negligible due to the much shorter average lifetime in this higher excitation energy range.

Cross-correlation fits to the solution of Eq. 2 are shown in Figs. 3A and 3B, yielding energy-averaged lifetimes of 31.8(7) fs for the nanorods and 29.9(8) fs for the nanoshells. In broad terms, this agreement suggests that the nanoparticle geometry in the tens-of-nanometer size range has little effect on the hot electron decay, which is instead primarily influenced by the nanoparticle material itself. Closer examination reveals several more specific implications: First, due to the large difference in surface-to-volume ratios for the nanorods ($S/V = 0.5$ nm$^{-1}$) and nanoshells ($S/V = 0.1$ nm$^{-1}$), it is evident that the surface must play a relatively minor role in the hot electron decay, at least in the absence of surrounding media with interfacial states/barriers in the relevant energy range. This is consistent with previous studies of hot carrier thermalization times in gold and silver nanosphere ensembles, which decreased appreciably from the metal film limit only for very small ($< 10$ nm) diameters (58). Second, charge transfer to the ITO substrate is evidently negligible in both systems, as this would accelerate the decay measured for the nanorods (with hot carrier density much closer to the substrate interface) relative to the nanoshells. Finally, the monocrystallinity of the nanorods (59) versus polycrystallinity of the nanoshells (60) has little effect, as expected from the quasi-elastic nature of electron-phonon/defect scattering.

A handful of nanorods with different SPRs are examined at constant pump photon energy (Fig. 3C), showing little evidence of detuning effects on the hot electron decay. In each case, the pump power is adjusted for consistent pump-probe count rates and therefore excited carrier densities. However, carrier density is also found to have no observable effect on the average lifetime, as measured on a single representative nanorod over a range of pump fluences (Fig. 3D) corresponding to calculated peak electron temperatures between 800–1700 K (see two-temperature modeling described in Supplementary Information). These results are in good agreement with the theory described below, which indicates a mere ~10% increase in the average lifetime from the zero temperature limit up to 1500 K (Fig. S6). This insensitivity to fluence and thus carrier density precludes any significant decay contributions from hot carrier-carrier scattering/recombination, the rates of which should depend on pump fluence and carrier density.

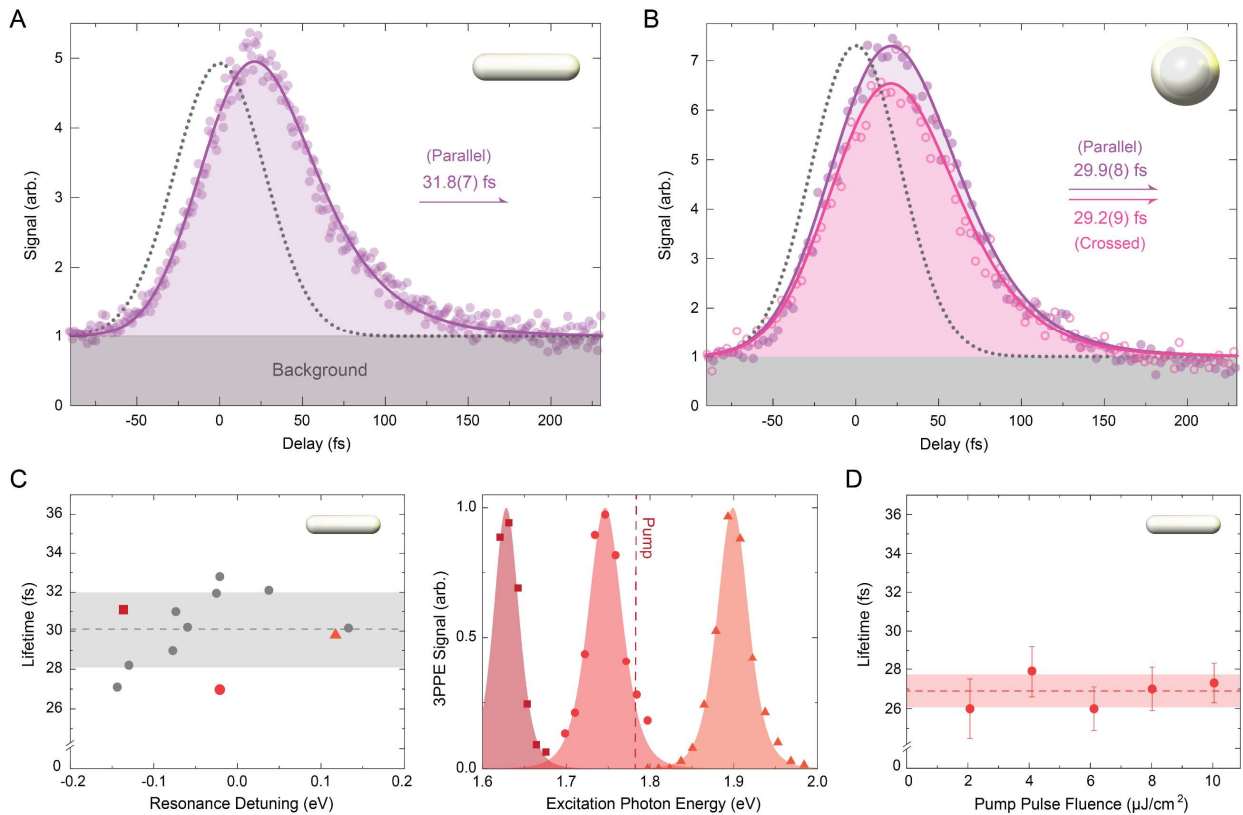

**Fig. 3. Energy-averaged femtosecond dynamics in single nanoparticles.** (**A**) Nanorod time delay scan fit to the instrument response function (dotted line) convolved with a forward exponential hot electron response. The instrument response function is the pump-probe cross-correlation measured on the ITO substrate with an unresolvable (< 10 fs) fast response time. Due to the strong longitudinal nanorod coupling of both pump and probe beams, data is collected with parallel-polarized pump and probe beams along this axis. (**B**) Nanoshell time delay scans and fits with both parallel- and cross-polarized pump and probe beams. (**C**) Summary of lifetimes with respect to longitudinal dipolar SPR detuning from the pump photon energy (1.77 eV) for 12 nanorods, with a mean value of 30.1 fs and a standard deviation of 1.9 fs. Single-color 3PPE spectra are shown for three representative nanorods, along with third-power Lorentzian fits. (**D**) Lifetimes are independent of incident power, as measured for a representative nanorod across the relevant pump pulse fluence range utilized in these studies.

For nanorods, parallel pump and probe polarizations along the longitudinal axis are necessary for measurable pump-probe signal due to the strongly enhanced interactions along this axis (Fig. 2D). By contrast, both parallel- and cross-polarized beam configurations can be utilized for the azimuthally symmetric gold nanoshells. We find that the total pump-probe signal in the cross-polarized configuration is reduced by only 14%, with a negligible effect on the energy-averaged hot electron lifetime (Fig. 3B). A unique feature of nanoplasmonic systems is the dramatically spatially-varying field profiles that depend strongly on incident optical parameters, including polarization and frequency. While expected for bulk and thin films, such a minor change in pump-probe signal for cross-polarized nanoshell excitation is noteworthy, providing information on the electric field profiles and bulk- versus surface-mediated excitation pathways. Specifically, simulations reveal that this strong cross-polarized pump-probe signal could only occur for bulk-like probe photon absorption, as the internal field is nearly isotropic for the probe excitation (Fig. S8). The probe surface field, by contrast, is strongest along the polarization axis and therefore well separated by distances greater than the ~40 nm inelastic mean free path from the pump excitation region, which lies along the pump polarization axis for either surface or bulk excitations (31) (Fig. S8). This strongly suggests that only bulk-like ballistic dynamics are probed within the nanoshells, as unambiguously confirmed below for nanorods via full velocity-resolved pump-probe studies.

**Velocity-resolved ballistic emission**

A two-dimensional (2D) projected velocity distribution for single nanorod excitation at the peak pump-probe time delay is shown in Fig. 4A, along with a central slice of the corresponding reconstructed 3D distribution. The 3D reconstruction is performed via inverse Abel transform using the Gaussian basis set expansion (BASEX) method (61), which relies on the approximate cylindrical symmetry in the photoemission distribution with respect to the longitudinal nanorod axis to compensate for information loss in the experimental 2D projection. Effects of deviations from perfect cylindrical symmetry for the ITO-supported nanorods are discussed in the Methods section. The full angle-integrated kinetic energy spectrum is then determined (Fig. 4C) from the reconstructed 3D distribution, with the background (i.e., non-pump-probe) contribution isolated at large time delays.

Even without 3D reconstruction, the 2D photoelectron velocity map offers considerable insight into nanorod hot carrier dynamics. Despite the longitudinal laser polarization, volume-mediated photoexcitation yields predominantly *transverse* ballistic hot electron escape from the nanorod side walls, leading to an orthogonal photoemission velocity distribution with respect to the longitudinal dipolar plasmon axis. In a recent study of gold nanorods, similar orthogonal emission behavior clarified the bulk-like nature of single-color multiphoton photoemission, which transitioned to a surface regime only with sufficient red detuning (32). This was demonstrated for nanorod resonances across the 2–4PPE regimes, with a greater tendency toward bulk-like excitation for lower-order processes due to a stronger dependence on the integrated volume field rather than the peak surface field. Bulk-like intraband excitation are therefore expected to be dominant for nanorods—as well as nanoshells (31) and other gold nanoparticle geometries—in the resonant one-photon absorption limit.

The observed orthogonal photoemission for the resonant + non-resonant pump-probe excitation here is consistent with the resonant pump absorption occurring predominantly throughout the field-enhanced central nanorod region (Fig. S8), although this may also occur due to rapid hot electron diffusion throughout the nanorod following surface-like excitation at the nanorod tip hot spots. Regardless, the photoelectron velocity distributions reveal that the non-resonant probe photon absorption occurs throughout the volume of the nanorod and, as a result, the probed dynamics are necessarily ballistic rather than interfacial in nature. Indeed, when combined with the observation of similar nanorod and nanoshell lifetimes, the surface appears to play little role beyond defining the boundary conditions for the plasmon modes and corresponding field distributions, even in nanoparticles with dimensions as small as 10 nm.

By way of theoretical comparison, the velocity-resolved photoemission distribution is calculated for a gold nanorod (Fig. 4B) using Monte Carlo sampling of ballistic final state electron trajectories, accounting for angle- and energy-dependent transmission at the surface potential barrier. We approximate rapid superdiffusive transport of the pump-excited hot electrons throughout the volume of the nanorod (elastic electron-phonon mean free path ∼30 nm; energy-averaged inelastic mean free path of ∼40 nm measured above in Fig. 3A for a gold Fermi velocity of 1.4 nm/fs in gold). The final state photoexcitation density is then proportional to the spatial distribution of $I_{probe}$ within the nanorod (Fig. 4B). The calculated 3D velocity slice shown in Fig. 4B exhibits similar transverse photoemission behavior to that observed experimentally. A nanorod work function of 4.15 eV is determined by best fit of a heated Fermi-Dirac distribution to the Fermi edge in the experimental kinetic energy spectrum (Fig. 4C). The hot electron energy distribution can be estimated as a Fermi-Dirac distribution, but is determined in further detail via kinetic Boltzmann theory at the peak delay time (see below and Supplementary Information). Further details of the Monte Carlo photoemission modeling are described in the Methods. While some disagreement is evident in the predicted and measured spectra at low kinetic energies, it may be noted that the experimental distribution lacks evidence of discrete energy levels that may occur in excitations mediated by interfacial states, and instead reiterates the expectation for bulk systems of a nearly uniform promotion of the ground state Fermi gas, as captured by the ballistic theory.

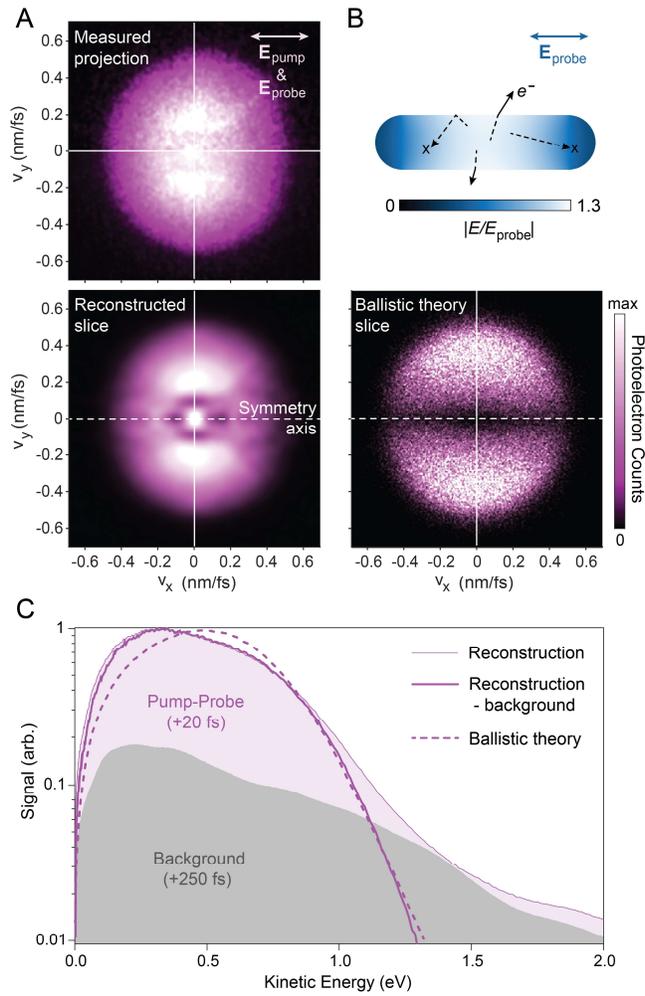

**Fig. 4. Pump-probe photoelectron velocity and kinetic energy distributions for a single gold nanorod.** (**A**) Measured projected velocity map and central $xy$ slice of the inverse-Abel reconstructed 3D distribution, with the axis of approximate cylindrical symmetry indicated. The zero-velocity peak is from ITO background counts. (**B**) Volume $xy$ slice of simulated internal field enhancement distribution under probe excitation, illustrating the Monte Carlo ballistic hot electron trajectory modeling, leading to the calculated velocity distribution slice shown beneath it. In both the measured and calculated distributions, the photoemission is predominantly *transverse* with respect to the nanorod longitudinal axis under longitudinal pump and probe polarizations. (**C**) Photoelectron kinetic energy distribution determined via angular integration of the reconstructed 3D distribution, along with the background-subtracted reconstructed distribution and the ballistic theoretical result.

**Energy-resolved femtosecond ballistic dynamics**

By combining both time and velocity resolution capabilities, the full energy-resolved femtosecond dynamics of hot electrons are measured in single resonantly-excited gold nanorods (Fig. 5A). As summarized in Fig. 5C, the lifetimes depend approximately inverse quadratically on the excitation energy, $E - E_\text{F}$. While this may appear characteristic of Fermi liquid behavior observed in nearly-free-electron systems such as aluminum and noble metals (20) (Supplementary Information), we offer a more rigorous analysis of the hot electron dynamics, also accounting for the femtosecond pulsed excitation, plasmon-enhanced superheating of the electron gas, electron-phonon coupling, and cascading/in-filling from higher energy levels. The temporal evolution of the hot electron distribution function, $f(E,t)$, is determined via the kinetic Boltzmann equation,

$$\frac{df(E,t)}{dt} = H(E,t) + \frac{df(E,t)}{dt}\bigg|_{e-e} + \frac{df(E,t)}{dt}\bigg|_{e-ph}, \tag{3}$$

in which $H(E,t)$ accounts for the time-dependent optical excitation of hot electrons, which proceed to relax via electron-electron and electron-phonon scattering (second and third terms, respectively). The electron-electron coupling is described by a screened Coulomb potential using the approximated Thomas-Fermi expression with a phenomenologically reduced screening wavevector $\beta k_\text{TF}$, such that $\beta = 1$ yields Thomas-Fermi screening while $\beta < 1$ corresponds to reduced screening and stronger electron-electron interactions (58). The electron-phonon coupling is modeled via interaction potential with an effective deformation potential constant. Absorbed energy densities consistent with experiments are utilized in the calculations, corresponding to a peak electron temperature of 1500 K as determined via two-temperature modeling (Supplementary Information).

The results of this analysis are shown in Fig. 5B and lifetimes are summarized in Fig. 5C, with a value of $\beta = 0.9$ yielding best agreement with experiment. The power-law exponent of the energy-dependent lifetimes predicted by kinetic Boltzmann theory is approximately $m_B = -2.4$, which is steeper than Fermi liquid theory ($m_{FLT} = -2.0$) due primarily to the influence of the longer-lived hot electron tail on the lower-energy states (Supplementary Information). In the zero-temperature limit (Fig. S6), $m_B \to -2.1$, still slightly elevated above the Fermi liquid theory value due to weak cascaded in-filling from higher to lower energy levels. This cascading is generally small for intermediate state energies greater than half the pump photon energy (> 0.9 eV here) (62). The effect of electron-phonon scattering is found to be negligible for the experimentally probed excitation energies, but quite significant for longer time scale relaxation kinetics in the < 1 eV range (Fig. S4). Experimentally, a power law fit with $m_{exp} = -2.3(1)$ is suggestive of the strong transient electron heating and corresponding deviations from Fermi liquid theory predicted by the full kinetic treatment, yet such discrepancies remain unclear within experimental uncertainty and require further investigation. With a single-parameter fit, the kinetic Boltzmann theory also provides an extrapolation of the experimental values across the full hot electron and hole distributions (Fig. S2), with much longer decays in in the low energy range that deviate considerably from single-exponential behavior due to more significant contributions from cascading, electron-phonon scattering, and thermalized electrons.

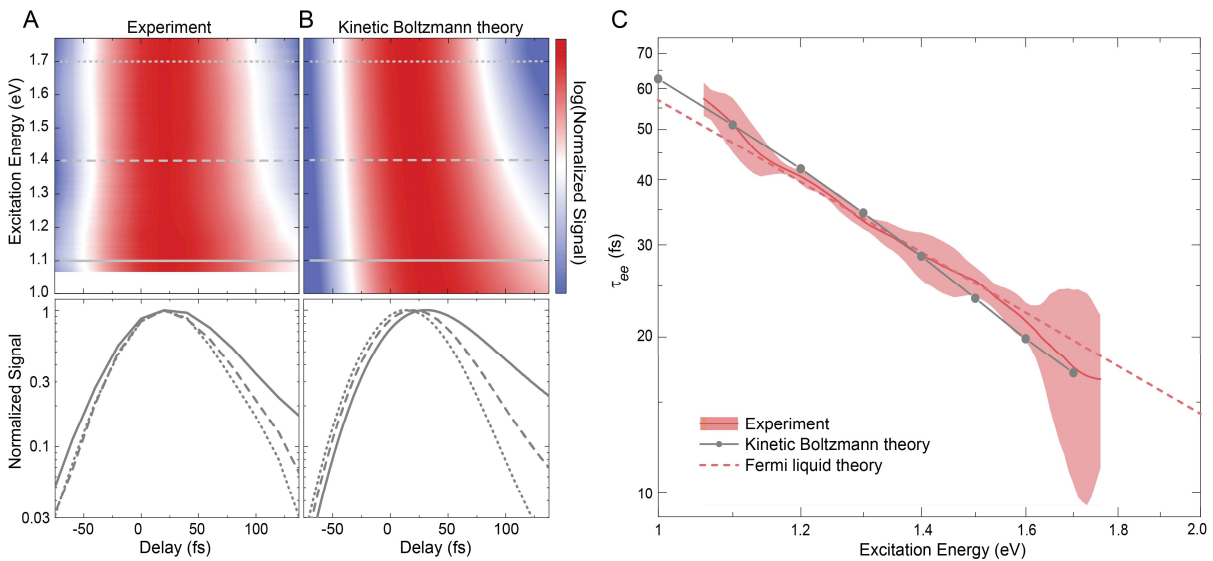

**Fig. 5. Full characterization of time- and energy-resolved hot electron dynamics in single gold nanorods.** (**A**) Experimental and (**B**) theoretical log-scale colormaps of the time-resolved pump-probe signal for a range of intermediate state excitation energies, $E_m = E - E_F$. The peak signal at each energy is normalized for clearer visual comparison of the dynamics. Cross-correlation time traces for three different excitation energy lineouts are shown beneath each plot. (**C**) Summary of hot electron lifetimes for the full measurable excitation energy range. Experimental average measured for four nanorods (shaded region is the standard deviation). The oscillations are attributed to integrated white noise from the velocity maps and vary from rod to rod. Kinetic Boltzmann theory with screening parameter $\beta = 0.9$ shown for comparison, along with the best Fermi liquid theory fit ($r_s = 1.87$; Supplementary Information).

## Discussion

The measured hot electron lifetimes in gold nanoparticles (Fig. 5C) lie between ~30% faster values measured previously for bulk gold (63) and ~50% slower values measured for thin gold films (20, 63, 64). Although measured lifetimes varied across different bulk and thin film studies, these general trends can be attributed to two competing effects that are avoided in the present study: transport and $d$-band excitations. Hot carrier transport out of the probing region leads to faster lifetimes measured in bulk metals compared with thin 10–25 nm films, where the hot carriers are constrained to a region around the escape depth (63, 64). Transport is similarly restricted in the nanoparticle limit, with hot carrier lifetimes indeed observed here to be longer than bulk gold values. Additionally, unlike in the preponderance of single-color femtosecond 2PPE studies of utilizing > 2 eV pump and probe photons, the plasmon-resonant pump photon energy (1.77 eV) used here falls below the ~2 eV $d$-band onset (Fig. 1A) and thus precludes interband absorption, instead featuring resonantly-enhanced intraband pumping and relatively weak probe fluences. The measured lifetimes are therefore unaltered by $d$-band effects—most prominently, the delayed Auger decay of $d$-band holes (20)—which we suggest leads to the faster lifetimes compared with previous thin film measurements. Most importantly, a wide variety of gold nanostructure geometries away from the spherical limit exhibit < 2 eV (> 620 nm) resonances that similarly minimize such $d$-band effects.

Other effects that have been shown to influence lifetimes in the nanoconfined limit are avoided here as well. In particular, electron wavefunction spillout and $d$-band related screening reduction in surface atomic layers leads to enhanced electron-electron interactions that reduce hot carrier thermalization times (relative to Au films) for < 10 nm nanoparticles (17, 58). However, the > 10 nm nanoparticle dimensions and nearly identical average lifetimes measured in nanorods and nanoshells suggest that such effects are negligible in the present studies. While quantum confinement in few-nanometer metal clusters may increase hot carrier lifetimes relative to thin metal films (21) due to a reduced density of states and therefore reduced electron-electron interactions (65, 66), the nanoparticles studied here are well beyond this size regime.

Thus, with internal ballistic dynamics verified for isolated gold nanorods and nanoshells, we find that these intermediate-nanoscale geometries provide a uniquely unadulterated view of the intrinsic electron-electron decay times in gold. This also suggests that such nanoparticles can serve as an excellent platforms for introducing nontrivial material interfaces and isolating their effect on the dynamics. Furthermore, the lifetimes measured here provide the first direct experimental benchmarks for energy-resolved femtosecond dynamics in nanoscale gold, which can be extended quantitatively to other gold geometries, while qualitative insights from this discussion can be applied to other noble metals. It is our hope that such insights may be utilized to improve upon the generally low (< 1%) ballistic hot carrier collection efficiencies observed to date, which represent a crucial challenge for the development of next-generation hot carrier optoelectronic and photocatalytic devices.

## Methods

**Sample Preparation**

The quality and dimensional statistics of commercially-purchased nanorod (Nanopartz Inc.) and nanoshell (nanoComposix) samples are verified via transmission electron microscopy (FEI Tecnai T12 SpiritBT, 100 kV, $LaB_6$ cathode) after drop-casting 15 μL of the aqueous dispersions onto a carbon-coated TEM grid for 5 minutes, removing the excess solution, and drying in air. Samples are then prepared for photoemission studies by spin-coating 50 μL aliquots at 1500 rpm onto ITO-coated (10 nm) borosilicate coverslips (< 1 nm RMS surface roughness). UV-ozone cleaning of the substrates prior to deposition ensures good wetting. Nanoparticle solutions are diluted prior to deposition to ensure appropriately dense surface coverages (~5 nanoparticles per $10 \times 10$ μm$^2$ area) with average particle separations much greater than the diffraction-limited beam spot. Samples are prepared under ambient conditions and loaded into vacuum immediately upon preparation. Nanoparticles are "cleaned" under vacuum prior to all studies via brief (~1 s) exposure to high-intensity (~1 GW·cm$^{-2}$) 400 nm light in an area scan, which has been shown to reversibly remove air contaminants and eliminate post-emission scattering that obscures the final velocity distribution (30, 67).

**Scanning Photoemission Imaging Microscopy**

Our scanning photoemission imaging microscopy (SPIM) technique combines ultrafast pump-probe excitation, scanning photoemission microscopy, and velocity map imaging photoemission spectroscopy, enabling single-nanoparticle time-, angle-, and energy-resolved photoemission studies. The second harmonic of a Ti:sapphire oscillator (KMLabs Swift, 75 MHz, ~50 fs pulses, 700–1000 nm tuning) pumps an optical parametric oscillator, with the tunable signal beam output (515–775 nm) serving in these experiments as the plasmon-coupled pump beam. The remainder of the second harmonic (400 nm, 3.1 eV) serves as the probe beam, with the relative delay time scanned with a motorized delay stage in the probe beam path, as illustrated in Fig. 1. The pump and probe pulse durations are both approximately 50 fs, as determined by frequency-resolved optical gating measurements at the laser output and by photoemission cross- and auto-correlation measurements directly at the sample position (on the ITO substrate). While highly tunable, the pump beam is maintained at 700 nm for the present pump-probe studies for (i) optimal plasmon-resonant coupling for both nanorods and nanoshells, and (ii) optimized dispersion compensation and pulse duration. The present measurements are not interferometrically stabilized and are thus phase-averaged. Group velocity dispersion is optimized to minimize pulse durations via dual-prism compensators in both pump and probe beam paths, with the single-color photoemission signal monitored as a function of prism pair separation and insertion into the beam path.

The two beams are focused down to overlapping diffraction-limited spots on an $xy$ scanning sample stage using a reflective 0.65 NA microscope objective, all under high vacuum conditions ($2 \times 10^{-7}$ Torr), with the pump-probe photoemission collected as a function of both sample position (Fig. 1 inset) and time delay. Dozens of nanoparticles can be interrogated simultaneously (within a series of scans) as a function of different incident laser parameters (intensity, frequency, and polarization) via large-area scans. Alternatively, single nanoparticles can be studied over the course of hours with little drift relative to the diffraction-limited laser spot (or for days with minor positional corrections). Larger (millimeter-scale) $xy$ positioning and $z$ focusing is achieved with piezoelectric motors.

Photoelectrons emitted from the sample are linearly mapped onto a phosphor-microchannel plate detector via the electrostatic lens illustrated schematically in Fig. 1, which consists of three copper electrodes: (i) The sample stage (repeller electrode) biased at −4500 V,

(ii) an "extractor" plate biased at $-3700$ V, and (iii) a final grounded electrode plate. This velocity map imaging lens configuration serves to linearly map photoelectrons from initial velocity $(v_x, v_y)$ onto detector position $(d_x, d_y)$, integrating over initial $v_z$ with negligible distortion (given the much larger 4.5 keV accelerated photoelectron energies compared with < 1 eV initial energies) and little sensitivity to initial position (68). The repeller bias is adjusted to optimize use of a 75 mm microchannel plate detector area while minimizing distortion due to electrode proximity for the outermost electron trajectories. The extractor plate bias is then optimized via ion trajectory simulations (*SIMION 8.0*) to achieve the initial-position-insensitive velocity-mapping condition (68). The system is calibrated via laser frequency-dependent photoemission studies and Fermi edge fits for thin gold film, as described along with further details and characterization of the system in previous work (67).

**Finite Element Simulations**

Finite element simulations are performed using the RF module in *COMSOL Multiphysics 6.0* using plane electromagnetic waves normally incident on ITO/glass-supported gold nanoparticles. Average nanorod and nanoshell ensemble dimensions are utilized to model the nanoparticles. Measured values from Johnson and Christy (69) are used for the gold dielectric function, while the ITO dielectric function was determined via ellipsometry. A perfectly-matched layer is utilized for domain truncation and due diligence is performed for convergence with respect to mesh, domain, and PML size. Total photoemission values are approximated by integrating the volume field enhancement raised to the $2n$ power for $n$PPE (thus approximating a position-independent escape coefficient). This approximation is suitable for estimating simple polarization-dependent behaviors (Figs. 2D and 2E). For ballistic photoemission calculations (see below), the simulated volume field enhancement data is exported on a fine uniform mesh. Simulated surface and volume fields for both nanorod and nanoshell geometries are shown in the Supplementary Information, Fig. S8.

**Inverse Abel Transform**

Approximate 3D photoelectron velocity distributions are reconstructed from the measured 2D distributions via inverse Abel transform (where the projection in $v_z$ constitutes the forward Abel transform). This is implemented efficiently via matrix algebra in *MATLAB* using the basis set expansion (BASEX) method of Dribinski *et al*. (61) with nearly-Gaussian basis functions. A standard deviation of $\sigma = 1$ px and regularization parameter of 10 are utilized.

The information lost in the 3D → 2D projection can be recovered if an axis of cylindrical symmetry exists orthogonal to the projection direction (i.e., within the $xy$ plane here), thereby permitting a unique reconstruction. The validity of such an approximation is expected to be perfect for unsupported cylindrical nanorods, though could be called into question for the nanorod faceting and ITO substrate. However, the typical octagonal nanorod side faceting (59) and nominal $\cos(\theta)$ photoelectron emission distribution with respect to the local surface normal (67) yields a quasi-continuous azimuthal distribution about the nanorod longitudinal axis. Furthermore, several points can be made on the symmetry-breaking effects of the ITO substrate: (i) The internal probe field, corresponding volume hot carrier excitation density, and emission distribution remain reasonably symmetric (25% top/bottom field intensity asymmetry), including further blurring out due to the internal ballistic dynamics. (ii) For the downward-moving photoelectron trajectories that aren't immediately beneath the nanorod (i.e., those that escape into free space), the projected velocity distribution would not be altered by angle/energy-independent specular reflection or collection by the ITO. Clearly, even basic considerations of the quantum reflection probability at a step-down barrier, $|R|^2 = |(k_\perp - k'_\perp)/(k_\perp + k'_\perp)|^2$, can be expected to distort the velocity distribution of the substrate-reflected photoelectrons and lead to deviations from perfect cylindrical symmetry. While more careful accounting of such effects may be valuable in future studies, the good agreement between the reconstructed experimental and calculated photoelectron distributions (Fig. 4) suggests that these deviations are small.

**Ballistic Dynamics Simulations**

Briefly, for pump-probe 2PPE, it is approximated that the pump-excited hot electrons (traveling at ≳1.4 nm/fs) are uniformly distributed within the nanorod when probe absorption occurs. Thus, only the internal probe field intensity distribution need be accounted for. A Monte Carlo method involving randomly-sampled trajectories for final-state electrons is then employed, taking into consideration (i) experimentally-determined energy-dependent inelastic mean-free paths within the nanorod (from measured lifetimes), (ii) angle-dependent quantum transmission coefficients at the metal-vacuum interface (depending on the surface-normal momentum), and (iii) "refraction" of escaped photoelectrons due to surface-normal momentum loss and parallel momentum conservation at the interfacial potential barrier. The geometry of the nanorod and the probe-intensity-weighted spatial excitation distribution (determined from finite element simulations) are fully accounted for in these calculations, which can be extended to arbitrary 3D geometries. Further details on this Monte Carlo hot electron emission modeling can be found in previous work (32).

**Kinetic Boltzmann Theory**

The relaxation dynamics is modeled using the quasiparticle approach of the Fermi-liquid theory and describing the conduction-electron system by a one-particle distribution function $f(E, t)$. Its time evolution is given by the Boltzmann equation (Eq. 3) (10, 16, 58, 70, 71). The initial out of equilibrium excitation by intraband absorption of a femtosecond pump pulse of frequency $\omega_{\text{pump}}$ (first term of Eq. 3) leads to the creation of a spatially homogeneous athermal distribution where conduction electrons with an initial state energy, $E_i$, between $E_F - \hbar\omega_{\text{pump}}$ and $E_F$ are excited above the Fermi energy to an intermediate state energy, $E_m$, between $E_F$ and $E_F + \hbar\omega_{\text{pump}}$.

Modification of $\Delta f$ induced by electron-electron scattering (second term in Eq. 3) is described by a screened Coulomb potential, with an $|\epsilon|^{-2}$ screening factor depending on interband and Drude-like electron dielectric functions $\epsilon$. In the limit of small wavevector and energy exchanges, $\epsilon \approx \epsilon_{\text{ib}}^0[1 + (\beta q_{\text{TF}})^2/q^2]$, where $\epsilon_{\text{ib}}^0$ is the long-wavelength static value of the interband (i.e., from $d$-band to conduction band) electron screening contribution, $q_{\text{TF}}$ is the Thomas-Fermi screening wavevector, $\beta$ is a phenomenological screening reduction parameter, and $q$ is the wavevector exchanged during an electron-electron collision. The screened Coulomb potential is then multiplied by electronic state occupation numbers to account for probability of transitions from occupied initial states to available final states.

The electron-phonon scattering contribution (third term in Eq. 3) describes electron energy loss by energy to lattice energy transfer. This is modeled starting from an interaction potential with an effective deformation potential constant and considering an isotropic band for phonon energies. Further details are provided in the Supplementary Information.

Numerical computation of these three contributions leads to the full solution of the Boltzmann equation for the time evolution of $f(E, t)$. For all computations in this work, the pump pulse is set to 1.8 eV phonon energy and 50 fs duration, with an excitation temperature (peak electron temperature increase given the injected pump pulse energy) of 1200 K (Fig. S3). Final $\Delta f$ values are obtained after energy and time convolution of the probe pulse with 0.1 eV energy width and 50 fs time duration.

## Author Information

**Notes**

The authors declare no competing interests.

**ORCID iDs**

Jacob Pettine: 0000-0003-2102-1743

Paolo Maioli: 0000-0002-4199-8810

Natalia Del Fatti: 0000-0002-8074-256X

David J. Nesbitt: 0000-0001-5365-1120

## Data Sharing

All presented data and code utilized in this manuscript is available from the corresponding authors upon reasonable request.

## Acknowledgements

This work was supported by the Air Force Office of Scientific Research (FA9550-15-1-0090) and the National Science Foundation Physics Frontier Center (PHY-1734006). J.P. acknowledges additional support by the Laboratory Directed Research and Development program of Los Alamos National Laboratory under project number 20210845PRD1. N.D.F acknowledges the Institut Universitaire de France (IUF).

**Supplementary Information for**

# Energy-Resolved Femtosecond Hot Electron Dynamics in Single Plasmonic Nanoparticles

Jacob Pettine,[1,2,3] * Paolo Maioli,[4] Fabrice Vallée,[4] Natalia Del Fatti,[4,5] and David J. Nesbitt[1,2,6] *

*Correspondence may be addressed to J.P. (jacob.pettine@lanl.gov) or D.J.N. (djn@jila.colorado.edu)*

**This Supplementary Information includes:**

Supplementary text
Figures S1 to S8
Legend for Movie S1
Supplementary References

**Electron temperature evolution after ultrafast excitation: Boltzmann equation**

Before excitation, the electron energy distribution in the conduction band is described by a standard Fermi-Dirac function,

$$f(E, t = -\infty) = \frac{1}{e^{(E-\mu)/k_B T_e} + 1}, \quad (S1)$$

where $T_e = T_0$ is the initial temperature, $k_B$ is the Boltzmann constant, and $\mu$ is the chemical potential (or Fermi level) for finite temperature, which is equivalent to the Fermi energy (5.53 eV for gold) at 0 K. The chemical potential deviates only by $\Delta\mu(T_e) = -E_F(\pi k_B T_e/2E_F)^2/3$ from the Fermi energy at elevated temperatures (1), changing by a mere 0.002 eV (0.04%) at $T_e = 1500$ K (the highest electron temperature reached in the present studies). The most comprehensive description of the system evolution is given by the full numerical solution of the kinetic Boltzmann equation (Eq. 3 of the main text), $f(E, t)$ being the time-dependent electron distribution function. The three terms of the kinetic Boltzmann equation (Eq. 3) describe respectively:

I. ultrafast optical excitation,
II. energy redistribution within the excited electron gas (internal electron thermalization by electron-electron interactions),
III. energy loss of the electron gas by electron-phonon interactions.

The time and energy dependence of $\Delta f(E, t) = f(E, t) - f(E, -\infty)$ is plotted in Figs. S1 and S2, and the complete time evolution is shown in Movie S1. For a more detailed description of the problem, we refer to supplementary references (2-5). Each step/term may be summarized as follows:

I.

Immediately following absorption, ultrafast excitation creates an athermal electron energy distribution, a class of electrons with initial energies in the range $E_F - \hbar\omega_p < E < E_F$ below the Fermi level are excited into states with energies $E_F < E < E_F + \hbar\omega_p$ above the Fermi level. The number of electrons excited is proportional to the total pump pulse energy absorbed by the system, which is given at time $t$ by

$$E_{abs}(t) = \int_{-\infty}^{t} \sigma_{abs} I_{pump}(\tau) d\tau, \quad (S2)$$

in which $\sigma_{abs}$ is the absorption optical cross-section of the nano-object (here a nanorod or a nanoshell) and $I_{pump}(t)$ is the pump pulse intensity envelope. This equation implies instantaneous plasmon dephasing, which is a permissible approximation considering that the plasmon dephasing time ($< 10$ fs (6)) is much faster than any of the hot electron kinetics in the energy range of interest. The question of interest here is then how this excess energy evolves within and outcouples from the electron gas.

II.

Initially, $\Delta f(E, t)$ extends over a very broad energy range, with a shape modified by electron relaxation even during the pulse duration (Figure S2, for 0 fs, defined as the pump pulse maximum). $\Delta f(E, t)$ subsequently strongly narrows as the electron gas internally thermalizes, the perturbed zone being eventually limited to a region of the order of $k_B T_e$ around $E_F$. This is a consequence of the very fast relaxation of electrons with energy far from $E_F$, with an energy-dependent *e–e* scattering rate close to the form $\tau(E)^{-1} \propto (E - E_F)^2$ (see Fermi Liquid Theory below). This process is described by a screened Coulomb interaction potential, containing a sum over all possible two-electron scattering processes satisfying energy and momentum conservation. Although a single electron–electron scattering process is very fast (tens-of-femtoseconds), many of them are required for establishing a thermalized electronic temperature. To quantitatively reproduce the internal electron thermalization time experimentally observed in previous optical time-resolved experiment on bulk silver and gold, a phenomenologically reduced static screening of the *e–e* Coulomb potential is introduced. It accounts for screening overestimation of the simple static Thomas-Fermi model, and it implies using an effective screening wavevector, $q_S$, instead of the Thomas-Fermi wavevector, $q_{TF}$, where $q_S = \beta q_{TF}$. A factor $\beta < 1$ corresponds to reduced screening and stronger electron–electron interactions (i.e., shorter electron state lifetimes) with respect to Thomas-Fermi model. Also note that, as discussed below, the electron relaxation dynamics depend on the excitation strength for strong excitations (peak $T_e \gg T_0$). This is a consequence of the dependence of the electron scattering rates on the electron distribution (2).

III.

Electron energy is transferred from the electron gas to the lattice through electron-phonon interactions. Their contribution to Boltzmann equation (third term in Eq. 3) is modeled using an interaction potential with an effective deformation potential constant and considering an isotropic band for phonon energies. The electron-phonon interaction amplitude is used as a parameter set by reproducing the energy transfer rate measured for long delays and weak perturbations (~1 ps time constant for bulk Au) (2, 5). This electron energy loss is reflected in the time decay of the electron distribution function, $f(E, t)$, which after a few picoseconds returns to the new equilibrium temperature of the electron-lattice system, which is slightly above $T_0$.

**Simplified picture after electron thermalization: Two-temperature model**

After internal electron thermalization (step II), the electron distribution is again described by a Fermi-Dirac distribution at temperature $T_e$. Electron dynamics is thus fully described by the evolution of $T_e$. Boltzmann equation leads to the well-known rate-equation system of the two-temperature model, which takes the form

$$C_e(T_e)\frac{dT_e}{dt} = -g(T_e - T_l), \tag{S3a}$$

$$C_l \frac{dT_l}{dt} = g(T_e - T_l). \tag{S3b}$$

where $T_l$ is the lattice temperature and $g = 2 \times 10^{16}$ W K$^{-1}$m$^{-3}$ is the utilized gold electron-phonon coupling constant per unit volume. The temperature-dependent free-electron Sommerfeld specific heat is $C_e(T_e) = \pi^2 k_B^2 T_e n_e/(2E_F)$ with conduction electron density $n_e = 5.9 \times 10^{22}$ cm$^{-3}$, and is thus proportional to electronic temperature $T_e$, while lattice specific heat is $C_l = 2.4 \times 10^6$ J K$^{-1}$m$^{-3}$. Spatial dependence of the electron temperature is neglected, as the mean free path for bulk electron-electron scattering is similar to or larger than the dimensions of the nanorods studied here ($l_{ee} = \tau_{ee} v_F$; tens of nanometers for tens-of-femtosecond lifetimes and Fermi velocity $v_F = 1.4$ nm/fs in gold). Due to the $C_e(T_e)$ dependence on $T_e$, electron temperature decay slows down for stronger excitation. As a result, strong excitation (large values of $E_{abs}$) induces modifications of both electron-electron scattering and electron-phonon interactions. Evolution of $T_e$ and $T_l$ according to both the solution of the Boltzmann equation (by computing an equivalent $T_e(t)$ and $T_l(t)$ at each time) and the simplified two-temperature model for the excitation considered in this work are plotted in Fig. S3.

**Density Functional Theory**

The electronic structure of gold is well known and recalculated here for visualization of the relevant energies and intraband 1 + 1' photoemission pathways (Fig. 2A). We use the *Quantum Espresso* distribution (7) with a plane wave basis set, full-relativistic ultrasoft pseudopotentials, PBEsol exchange-correlation functional, and a DFT + $U$ correction to the localized $d$-band energies ($U = 2.0$ eV based on experimental fitting (8)). The 4.078 Å lattice constant of gold is utilized and verified for these calculations via self-consistent field optimization of the total energy. A $12 \times 12 \times 12$ $k$-point grid is utilized, along with a 64 hartree wavefunction cutoff energy, determined via convergence of the total energy in self-consistent field calculations.

**Fermi liquid theory**

Based on the many-body theory of Quinn for a free-electron gas (9), the Fermi liquid theory lifetime can be written (10) in terms of the effective electron density parameter, $r_s$ (or the effective electron density, $n_{e,\text{eff}} = \frac{3}{4\pi}(a_0 r_s)^{-3}$; with Bohr radius, $a_0$), as

$$\tau_{ee}(E_m) = \frac{2}{3}\left(\frac{3}{2\pi}\right)^{8/3} \frac{e^4 m_e}{\epsilon_0^2 \hbar} r_s^{-5/2} (E_m - E_F)^{-2}, \tag{S4}$$

in which $-e$ is the electron charge, $m_e$ the free electron mass, $\epsilon_0$ the permittivity of free space, $\hbar$ is Planck's reduced constant, and $E_F = 5.53$ eV is the Fermi level of gold, where we note that these "effective" $r_s$ and corresponding $n_{e,\text{eff}}$ parameters include $d$-band screening (11). This inverse quadratic lifetime dependence on the excitation energy is also readily derived from Fermi's golden rule by assuming approximately constant joint density of states (valid for gold around the Fermi level (12)) and transition matrix elements. This is known as the random-$k$ approximation (13-15), which has been shown to be well-justified for the coinage metals (Cu,

Ag, and Au) (10). Notably, the best fit value to our data in Fig. 5C is given by $r_s = 1.87$, which is 15% higher than the value ($r_s = 1.65$) determined experimentally for gold films (16). This corresponds to 40% lower lifetimes for the present nanoparticle measurements compared with previous thin film measurements, as discussed in the main text.

**Extraction of lifetimes from the kinetic Boltzmann equation solutions**

The simulated time-dependent $\Delta f(E, t)$ signals in the experimental energy range $\approx$ 1–1.8 eV (Fig. 5B in the main text and Fig. S4) feature two decays: a short one (20–100 fs) due to hot-cold electron-electron interactions, and a longer one (~few ps) due to electron-to-lattice energy transfer by electron-phonon scattering. The shorter lifetimes are the main focus in this work and are compared to those measured by energy- and time-resolved photoemission. Note that the influence of longer dynamics is negligible for large energies (Fig. S4), but not for energies closer to $E_F$ ($E - E_F < 1$ eV), for which a rise time is also present in the signal, as discussed below. To extract the energy-dependent hot electron lifetimes from numerical solutions of the kinetic Boltzmann equation, we fitted the simulated time-dependent $\Delta f(E, t)$ at constant $E$ with a fit function given by the analytical convolution of (i) a physical response function,

$$\Theta(t) * \left( A\, e^{-t/\tau_{ee}(E)} + B\, e^{-t/\tau_{ep}(E)} \right), \qquad (S5)$$

and (ii) a function accounting for the cross-correlation of pump and probe duration,

$$\frac{2}{\Delta t} \sqrt{\frac{\ln(2)}{\pi}}\, e^{-4\ln(2)\frac{t^2}{\Delta t^2}}, \qquad (S6)$$

where $\Theta(t)$ is the Heaviside step function, $\tau_{ee}(E)$ is the energy-dependent electron lifetime due to electron-electron scattering, $\tau_{ep}(E)$ the long-delay relaxation time due to electron-phonon interactions, $A$ and $B$ are amplitudes and $\Delta t$ is the pump-probe cross-correlation duration ($50\sqrt{2}$ fs).

An example of a numerical signal and its fit is given in Fig. S5A for $E - E_F = 1.0$ eV. The electron-electron coupling is dominant, but a weak hundreds-of-femtosecond tail is nevertheless observed. The values extracted by the full fit with the two decaying functions are indicated. The lifetimes deduced by numerical solutions of kinetic Boltzmann equation are consistent with respect to different choices of parameters used in the computations and lifetime extraction. In particular, $\Delta f(E, t)$ dynamics may be simulated starting from the Boltzmann equation by switching off the term responsible for electron-phonon thermalization. In the absence of electron-phonon thermalization, $\Delta f(E, t)$ decreases to a flat background (after fast electron decay), instead of slowly decaying to 0 (Fig. S5B). A fit of this numerical signal with a mono-exponential decaying function and a constant vertical offset (convolved with the pump-probe cross correlation) yields values of electron lifetimes within 7%.

**Effect of the excitation temperature on the extracted lifetimes**

As shown in Fig. S6, hot electron lifetimes depend on the excitation strength. A ~20% increase in the lifetimes is expected for relatively high (1200 K) peak electron temperature increases compared with near-zero energy input (10 K peak electron temperature increase). A correct estimation of $E_{\text{abs}}$ (Eq. S2) and thus both $\sigma_{\text{abs}}$ and $I_{\text{pump}}$ is required for a careful comparison of experimental and theoretical predicted lifetimes (17). While a peak value of $\sigma_{\text{abs}} \approx$ 5000 nm$^2$ is calculated via finite element simulations for nanorods excited around their longitudinal plasmon resonance, this neglects additional plasmon damping due to ligands, defects, and other nonidealities of the system, as well as the effect of detuning from resonance. In comparing the average data obtained experimentally for different nanorods to results of simulations, we thus utilize an average value of $\sigma_{\text{abs}} = 2500$ nm$^2$ for these absorbed energy estimations, along with the experimentally-determined pulse duration (50 fs) and peak pulse intensity (1.8 × 10$^8$ W cm$^{-2}$; pulse fluence of 9 µJ·cm$^{-2}$). This corresponds to a peak electron temperature rise $\Delta T_e = 1200$ K, used in the Boltzmann calculations and Monte Carlo simulations. Note that in the weak perturbation regime (peak $\Delta T_e = 10$ K), an $n \approx -2.1$ power law for the energy dependence of the electron lifetime approaches Fermi liquid theory but remains slightly modified by cascading electrons.

As is clear from Fig. S4, and more generally from the full energy- and time-dependent behaviors shown in Fig. S1 down to 0.5 eV, the lower-energy kinetics (particularly below ~0.75 eV) cannot be fit to exponential decays due to (i) in-filling effects from higher energy levels, which mainly delay $\Delta f(E, t)$ rise, and (ii) increasing population at long times due to the thermalized electron tail, which slowly decreases following electron-to-lattice energy transfer. With the considerations described in the previous sections and in the main text, the Boltzmann theory calculations provide a valuable means of understanding the various contributions to energy-dependent electron decay (especially in-filling, temperature, electron-phonon coupling, and screening), while also allowing for extension of the experimentally observed values to lower energies. It is important to mention that optical time-resolved experiments (2) are primarily sensitive to the rise dynamics of electronic states very close to Fermi level (a few tens of meV) and showed a decrease of the thermalization time with excitation temperature. This is associated with the $\Delta f(E, t)$ increase of such low energy states, which is faster for stronger excitation. Time-resolved photoemission measurements (this work), on the other hand, investigate the ultrafast decrease of the population of electronic states more distant to $E_{\text{F}}$, which slows down for stronger excitation. All these effects are reproduced by the kinetic Boltzmann equation numerical solution.

**Effect of the screening corrections on the extracted lifetimes**

For the numerical calculations, only the absorbed pump pulse energy and pulse duration are needed. The $\beta$ parameter that modifies the Thomas-Fermi screening serves as a single fit parameter to achieve best agreement between the kinetic Boltzmann theory and measured energy-dependent lifetimes, as shown in Fig. S7. Note that uncertainties in the absorbed energy due to uncertainty in $\sigma_{\text{abs}}$ will also induce an error in the determination of the value of $\beta$ giving the best fit. By including all the sources of error, we deduce an optimal value of $\beta = 0.90 \pm 0.05$. This is in reasonable agreement with previous measurements of time-resolved photoemission in other metals (i.e., bulk Ag (2)), where lifetimes could be reproduced using $\beta \approx 0.75$ with however very large uncertainties.

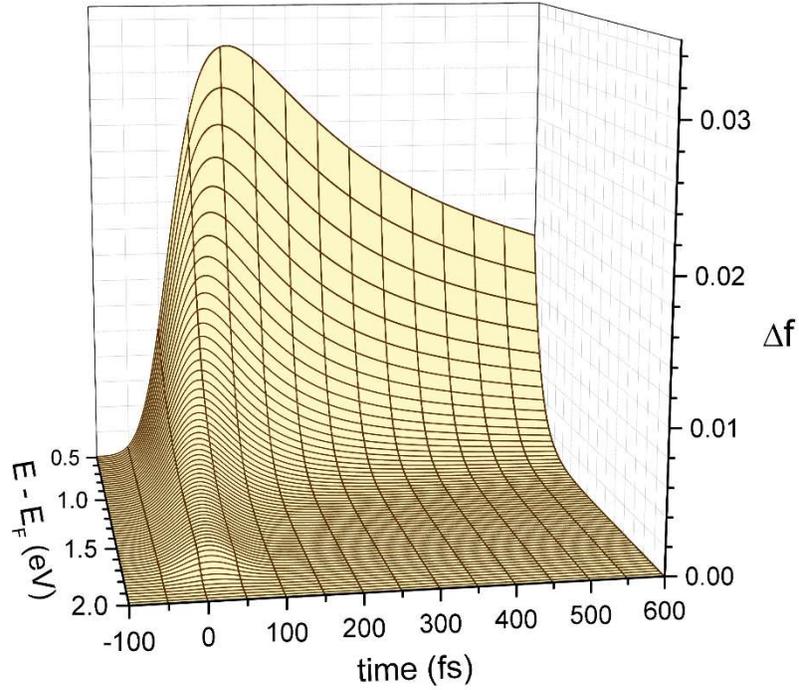

**Fig. S1.** Electron energy distribution variation, $\Delta f(E,t) = f(E,t) - f(E,-\infty)$, computed using the Boltzmann equation as a function of time delay after initial excitation and of electron energy with respect to Fermi level, $E_{\rm F}$. Much larger populations (and slight positive time shifts) at lower energies are due primarily to in-filling effects from electrons scattered out of higher energy states during internal electron thermalization. A pump photon energy of 1.8 eV and pulse duration of 50 fs are utilized. The screening reduction parameter is $\beta = 0.9$ and the absorbed energy corresponds to a peak $T_e = 1500$ K.

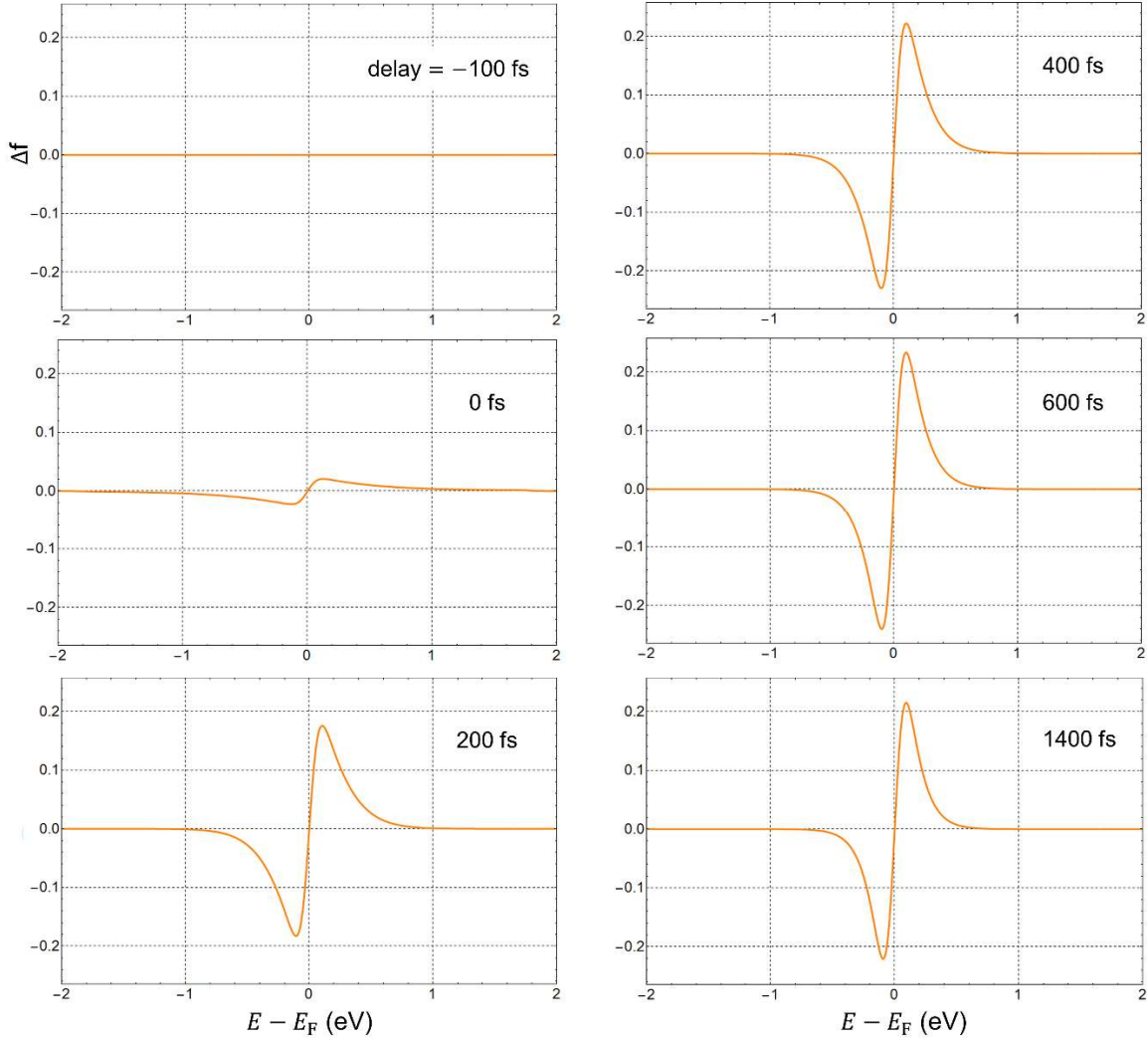

**Fig. S2.** Snapshots of Movie S1, showing the full hot electron (and hot hole) distribution function variation, $\Delta f(E,t) = f(E,t) - f(E,-\infty)$, as a function of electron energy. $\Delta f(E,t)$ starts from $-100$ fs (i.e., before optical excitation), then it increases (decreases) for energies above (below) Fermi level. This corresponds to pump-induced electron state filling (emptying), induced by electron transitions from occupied states below $E_F$ to unoccupied states above $E_F$ concomitant with the absorption of one pump photon. Through electron-electron interactions, excess energy is progressively transferred to states closer to $E_F$ on 200 to 600 fs timescales, before significant energy transfer occurs out of the electron gas by electron-phonon interactions ($> 600$ fs).

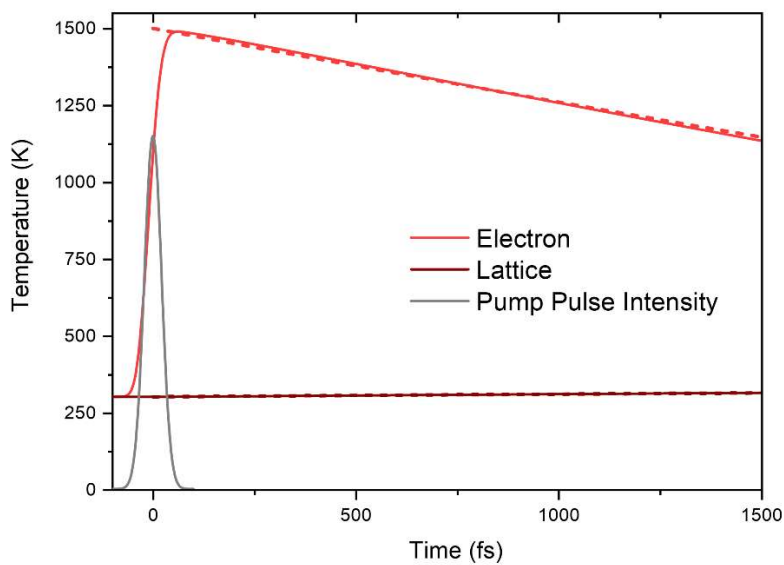

**Fig. S3.** Evolution of electron temperature, extracted from Boltzmann equation (solid red line) and from two-temperature model (dashed red line), and of lattice temperature (blue lines), after pump excitation corresponding to peak $T_e = 1500$ K, shown along with the pump pulse intensity envelope. The effective electron temperature at short times is computed as the temperature equivalent to the one of a thermalized electron distribution with same energy.

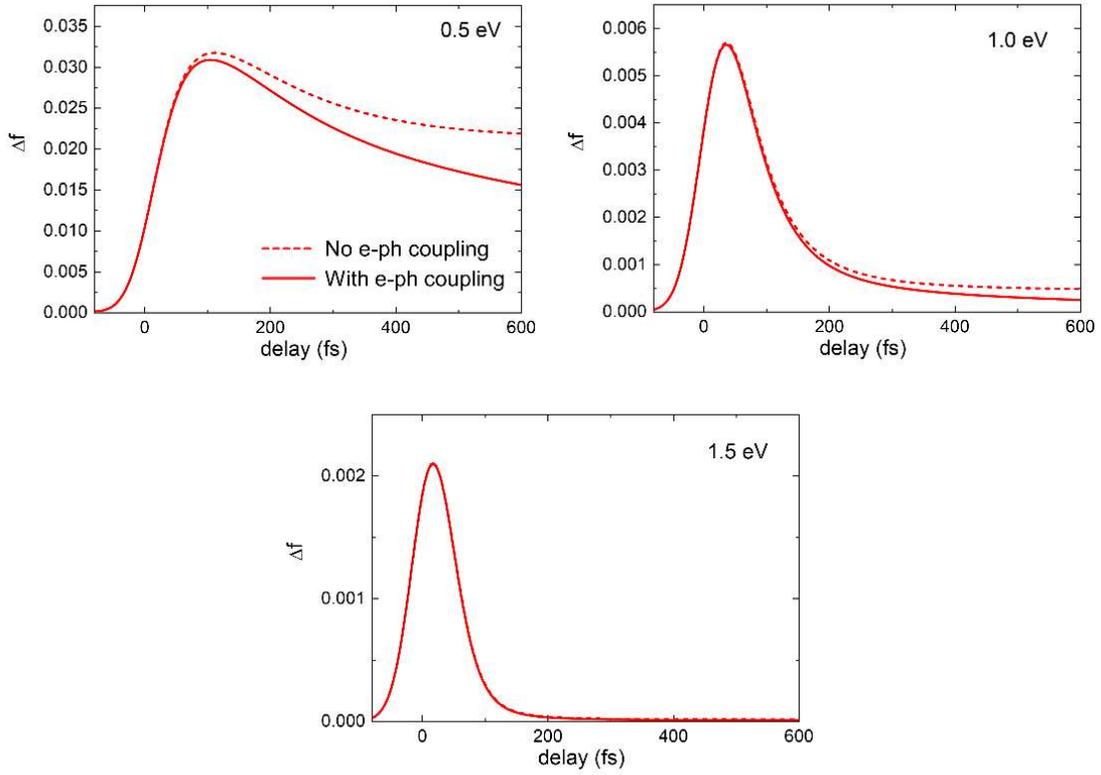

**Fig. S4.** Electron distribution variation, $\Delta f(E,t) = f(E,t) - f(E,-\infty)$, computed using the Boltzmann equation as a function of time for different electron energy states ($E - E_\mathrm{F} = 0.5$, 1.0 and 1.5 eV). The effect of electron-phonon coupling (solid lines) or the absence thereof (dashed lines) on the electron population kinetics is simulated at different energies. The effect is minimal for the experimental > 1 eV range (but becomes significant for lower energies). Same numerical parameters as in previous figures.

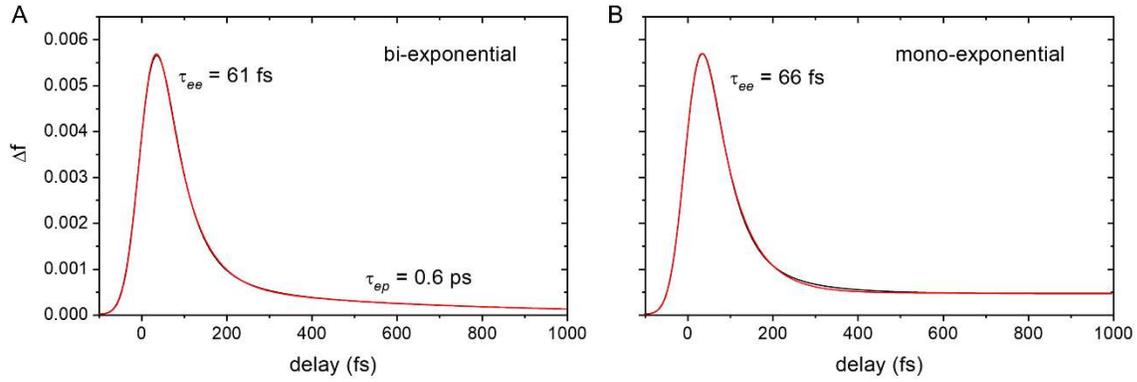

**Fig. S5.** (**A**) Electron distribution function variation $\Delta f(E, t)$ (black line) computed for $E - E_\mathrm{F} = 1.0$ eV and bi-exponential fit (red) accounting for both femtosecond and picosecond relaxation decays. (**B**) $\Delta f(E, t)$ computed with no e-ph interaction term and mono-exponential fit with offset for positive delays.

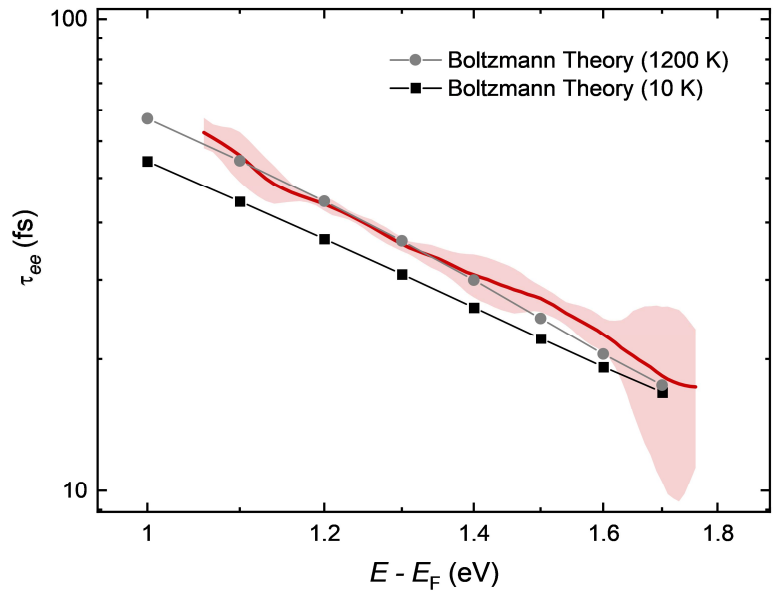

**Fig. S6.** Effect of pump pulse fluence and corresponding peak electron temperature increases $\Delta T_e = 10$ K versus 1200 K on the energy-dependent lifetimes, obtained by numerical solution of the kinetic Boltzmann equation. The value $\Delta T_e = 1200$ K (or peak $T_e = 1500$ K) corresponds to the approximate experimental value utilized for Monte Carlo modeling. The screening reduction parameter is $\beta = 0.9$. A pump photon energy of 1.8 eV and pulse duration of 50 fs are utilized.

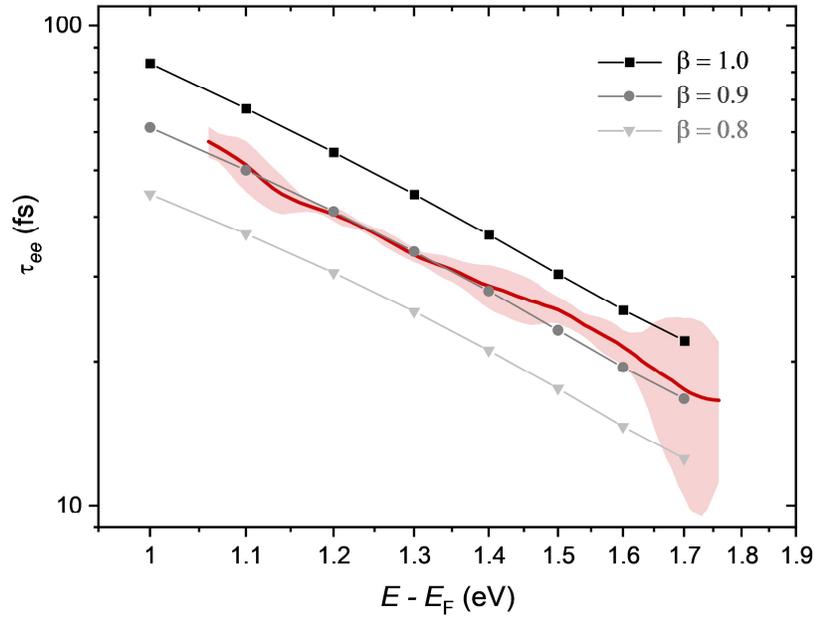

**Fig. S7.** Dependence of lifetimes on the $\beta$ parameter including correction for Thomas-Fermi screening. Lower $\beta$ values correspond to a weaker screening, which translates into stronger electron-electron interactions and shorter excited state lifetimes. Assuming an absorbed energy corresponding to a peak $T_e = 1500$ K, the value $\beta = 0.9$ is determined to yield the best fit and is utilized for all other calculations.

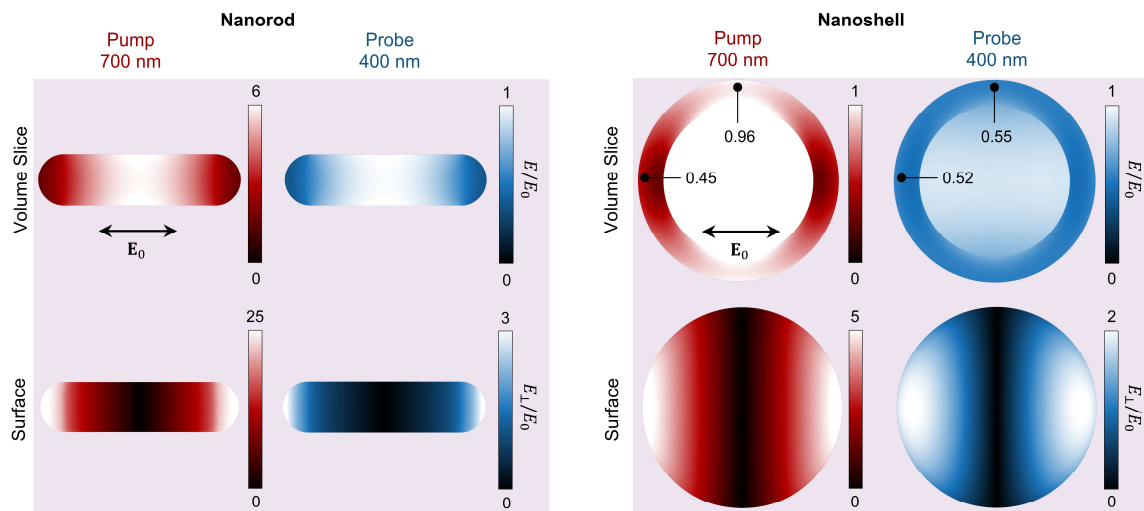

**Fig. S8.** Calculated plasmonic field distributions for the gold nanorods and nanoshells at resonant pump and non-resonant probe frequencies. The relevant surface-normal fields for surface excitation and total volume fields for internal excitation are shown. (Left) The internal nanorod field is highly centralized compared with the uniform field of an ellipsoid, due to the hemispherical charge buildup and local cancellation within the tip regions. (Right) The volume nanoshell field intensity is highly isotropic for the probe excitation, while all other volume and surface fields are highly anisotropic. Thus, the similar signals for parallel- and cross-polarized pump-probe time traces in Fig. 3B can only be rationalized in terms of a bulk-like (rather than surface) probe excitation.

## Supplementary References